\documentclass{ws_procs9x6}

\begin{document}

\title{Dissipative Chaos in Quantum Distributions}

\author{T. V. GEVORGYAN $^{1,*}$, S. B. MANVELYAN $^{1}$}
\address{$^{1}$Institute for Physical Researches,
National Academy of Sciences,\\Ashtarak-2, 0203, Ashtarak,
Armenia\\
$^*$E-mail: t\_gevorgyan@ysu.am\\
www.ipr.sci.am}

\author{A. R. SHAHINYAN$^{2, **}$, G. Yu. KRYUCHKYAN$^{1,2,***}$}
\address{ $^{2}$Yerevan State University, A. Manoogian 1, 0025,
Yerevan, Armenia \\
$^{**}$E-mail: anna\_shahinyan@ysu.am,
$^{***}$E-mail: kryuchkyan@ysu.am \\
www.ysu.am}

\begin{abstract}
We discuss some problems of dissipative chaos for open quantum
systems in the framework of semiclassical and quantum
distributions. For this goal, we propose a driven nonlinear
oscillator with time-dependent coefficients, i.e. with
time-dependent Kerr-nonlinearity and time-modulated driving field.
This model showing both regular and chaotic dynamics in the
classical limit is realized in several experimental schemes.
Quantum dissipative chaos is analyzed on the base of numerical
method of quantum trajectories. Three quantities are studied: the
Wigner function of oscillatory mode from the point of view of
quantum-assemble theory and both semiclassical Poincar$\acute{e}$
section and  quantum Poincar$\acute{e}$ section calculated on a
single quantum trajectory. The comparatively analysis of these
distributions for various operational chaotic regimes of the
models is performed, as well as scaling invariance in dissipative
chaos and quantum interference effects assisted by chaos are
discussed.
\end{abstract}

\keywords{Quantum optics, Nonlinear dynamics}

\bodymatter

\section{Introduction}\label{sec:intro}
Quantum nonlinear systems with chaotic classical counterparts have
received much attention in the last two decades. This field of
investigation is sometimes called quantum chaos\cite{quant_chaos}.
The majority of studies of quantum chaos for isolated or so-called
Hamiltonian systems, the classical counterparts of which are
chaotic, focus on static properties such as spectral statistics of
energy levels and transition probabilities between eigenstates of
the system. A variety of studies have also been carried out to
understand the features of time-dependent chaotic systems. In
contrast to that very little work has been done to investigate
quantum chaos for open nonlinear systems. The beginning of study
of an open chaotic system can be dated back to the
papers\cite{back} where the authors have analyzed the kicked rotor
and similar systems with discrete time interacting with a heat
bath. Quite generally, chaos in classical conservative and
dissipative systems with noise has completely different
properties, e.g., strange attractors can appear only in
dissipative systems. For dissipative systems Poincar$\acute{e}$
section has form of strange attractor in phase space while for
Hamiltonian systems it has form of close contours with
separatrices.

In classical mechanics a standard characterization of chaos might
be given in terms of the unpredictability of phase-space
trajectories or Poincar$\acute{e}$ section that consists of
non-localized distributed points. However, the most important
characteristic of classical chaotic systems-exponential divergence
of trajectories, starting at arbitrarily close initial points in
phase space — does not have quantum counterpart because of the
Hiesenberg uncertainty principle. The question has been posed of
what constitutes the quantum mechanical equivalent of chaos. Many
criteria have been suggested to define chaos in quantum systems,
varying in their emphasis and domain of application\cite{quant_chaos, lyap}. As yet,
there is no universally accepted definition of quantum chaos.

In recent years much effort has been expended, both theoretically
and experimentally, to explore the role of quantum fluctuations
and noise in the order-to-chaos transition for open systems. It is
obvious that the investigations in this area are connected with
the quantum-classical correspondence problem, in general, and with
the environment induced decoherence and dissipation, in
particular. Recently, this topic has been the focus of theoretical
investigations. As a part of these studies, it has been recognized
\cite{zur} that the decoherence has rather unique properties for systems
classical analogs of which are chaotic. In particular, the
formation of sub-Planck structure in phase space has been
discussed for chaotic system\cite{zur_2}. The connection between quantum
and classical treatments of chaos was also realized by means of
comparison between strange attractors on the classical Poincar$\acute{e}$
section and the contour plots of the Wigner functions\cite{X2}.

In this paper we investigate some problems of dissipative chaos
for open quantum systems in the framework of quantum
distributions. We use the traditional ensemble description of
Markovian open systems, based on the master equation. This
equation is presented in quantum trajectories in the framework of
the quantum state diffusion approach\cite{QSD}. Recently, it was
shown how quantum state diffusion can be used to model dissipative
chaotic systems on individual quantum trajectories\cite{spil,
perc}. In contrast with these papers, here we show that it is
possible to describe quantum chaos using also a statistical
ensemble of trajectories, which is actually realized in nature.

The requirement in realization of this study is to have a proper
quantum model showing both regular and chaotic dynamics in the
classical limit. We propose a driven nonlinear oscillator with
time-dependent coefficients for this goal. This model proposed to
study the quantum chaos in the series of papers \cite{SR, X1, X2,
X3, mer} allow us to examine challenging problems of quantum
dissipative chaos, including the problem of the quantum
counterpart of a strange attractor.

The other problem of our interest in this paper relates to the
quantum effects in systems with chaotic dynamics. Particularly, in
a recent paper, sub-Poissonian statistics of oscillatory
excitations numbers was established for chaotic dynamics of
nonlinear oscillator\cite{X1}. It was shown that
quantum-interference phenomena can be realized for the dissipative
nonlinear systems exhibiting hysteresis-cycle behavior and quantum
chaos\cite{X2, X3, mer}. The study of these phenomena provides a
fundamental understanding of quantum fluctuations in quantum chaos
and opens a way for new experimental studies of the quantum
dissipative chaos in the field of quantum optics.

The outline of this paper is as follows. In Sec.II we describe
both the model and the method of calculations. In Sec. III we
analyze correspondence between Poincar$\acute{e}$ sections and
Wigner functions for the chaotic dynamics. In Sec. IV we analyze
scaling invariance for quantum system and discuss quantum chaos
for the regimes in which the classical chaos is lost. In Sec.V we
present results on quantum interference phenomena for chaotic
dynamics. We summarize our results in Sec.VI.

\section{Driven nonlinear oscillator as an open quantum system}
In this section the systems and the methods of calculations are presented.
We treat the Duffing oscillator as an open quantum system and
assume that its time evolution is described by Markovian dynamics
in terms of the Lindblad master equation for the reduced density
matrix $\rho$. In the interaction picture that corresponds to the
transformation $\rho\longrightarrow e^{-i\omega a^{+}at} \rho
e^{i\omega a^{+}at}$, where $a^{+}$ and $a$ are the Bose
annihilation and creation operators of the oscillatory mode and
$\omega$ is the driving frequency, this equation reads as
\begin {eqnarray}
\frac{d\rho}{dt}= \frac{-i}{\hbar}[H_0+H_{int}, \rho]+\nonumber~~~~~~~~~~~~~~~~~~~~~~~~~~~~~~~~\\
\sum_{i=1,2}\left( L_{i}\rho
L_{i}^{+}-\frac{1}{2}L_{i}^{+}L_{i}\rho-\frac{1}{2}\rho L_{i}^{+}
L_{i}\right).\label{eq:master_equatin}
\end {eqnarray}
The Hamiltonians are
\begin{eqnarray}
H_{0}=\hbar\Delta a^{+}a, \nonumber ~~~~~~~~~~~~~~~~~~~~~~~~~~~~~~~~~~~~\\
H_{int}=\hbar\chi(t) (a^{+}a)^{2}+\hbar(f(t)a^{+}+
f(t)^{*}a)\label{eq:peturbation},
\end{eqnarray}
where $\chi(t)$ and $f(t)$, which may or may not depend on time,
represent, respectively, the strength of the nonlinearity and
amplitude of the force, $\omega_0$ is the resonant frequency,
$\Delta=\omega_0-\omega$ is the detuning. The dissipative and
decoherence effects, losses, and thermal noise are included in the
last part of the master equation, where $L_{i}$ are the Lindblad
operators:
\begin {equation} \label{eq:linbland}
L_{1}=\sqrt{(N+1)\gamma}a, \qquad L_{2}=\sqrt{N\gamma}a^{+},
\end{equation}
 $\gamma$ is the spontaneous decay rate of the dissipation
process and $N$ denotes the mean number of quanta of a heat bath.
Here we focus on the pure quantum effects and assume $N=0$.

This model seems experimentally feasible and can be realized in
several experimental schemes. In fact, a single mode e.m. field is well
described in terms of an anharmonic oscillator (AHO), and the nonlinear medium could be an
optical fiber or a $\chi(3)$ crystal, placed in a cavity. The
anharmonicity of mode dynamics comes from the self-phase
modulation due to the photon-photon interaction in the $\chi(3)$
medium. In this case, it is possible to realize time modulation of
the strength of the nonlinearity by using a media with periodic
variation of the $\chi(3)$ susceptibility.

On the other side, the Hamiltonian described by Eq.
(\ref{eq:peturbation}) describes a single nanomechanical resonator
with $a^{+}$ and $a$ raising and lowering operators related to the
position and momentum operators of a mode quantum motion
\begin {equation}
x=\sqrt{\frac{\hbar}{2m\omega_0}}(a+a^{+}),\nonumber \
p=-i\sqrt{2\hbar m\omega_0}(a-a^{+}) \label{eq:xp},
\end{equation}
where $m$ is the effective mass of the nanomechanical resonator,
$\omega_0$ is the linear resonator frequency and $\chi$
proportional to the Duffing nonlinearity. One of the variants of
nano-oscillators is based on a double-clamped platinum beam
\cite{N} for which the nonlinearity parameter equals to
$\chi=\hbar/4\sqrt{3}Qma_{c}^{2}$, where $a_c$ is the critical
amplitude at which the resonance amplitude has an infinite slope
as a function of the driving frequency, $Q$ is the mechanical
quality factor of the resonator. In this case, the giant
nonlinearity $\chi\cong3.4\cdot10^{-4} s^{-1}$ was realized. Note,
that details of this resonator, including the expression for the
parameter $a_{c}$, are presented in\cite{amp}. On decreasing
nanomechanical resonator mass, its resonance frequency increases,
exceeding 1 GHz in recent experiments\cite{nano, nano_1}. It is
possible to reach a quantum regime for such frequencies, i.e., to
cool down the temperatures for which thermal energy will be
comparable to the energy of oscillatory quanta. The recent
investigations in this direction are devoted to classical to
quantum transition of a driven nanomechanical
oscillator\cite{Qnr1}, generation of Fock states\cite{Qnr2},
nonlinear dynamics, and stochastic resonance\cite{Qnr3}. Note,
that in the last few years there has been rapid progress in the
construction and manipulation of such nanomechanical oscillators
with giant ${\chi}(3)$-Kerr nonlinearity. The nanomechanical
resonator with a significant fourth-order nonlinearity in the
elastic potential energy has been experimentally demonstrated
\cite{A}. It has also been shown that this system is dynamically
equivalent to the Duffing oscillator with varied driving force
\cite{duffing_force}. This scheme is widely employed for a large
variety of applications as well as the other schemes of micro- and
nanomechanical oscillators, more commonly as sensors or actuators
in integrated electrical, optical, and optoelectrical
systems\cite{nano, devices}.

Cyclotron oscillations of a single electron in a Penning trap with
a magnetic field are another realization of the quantum version of
the Duffing oscillator\cite{a,b,c}. In this case the anharmonicity
comes from the nonlinear effect that is caused by the relativistic
motion of an electron in a trap, while the dissipation effects
arise from the spontaneous emission of the synchrotron radiation
and thermal fluctuations of the cyclotron motion. Note that a
one-electron oscillator allows one to achieve a relatively strong
cubic nonlinearity, $\chi/\gamma \lesssim 1$.

In recent years the study of quantum dynamics of oscillators with
time-dependent parameters has been focus of considerable
attention. This interest is justified by many applications in
different contexts. Particulary, one application concerns to the
center of mass motion of a laser cooled and trapped ion in a Paul
trap\cite{trapped_ion}. The quantum dynamics of an AHO with time
dependent modulation of its frequency and nonlinearity parameters
has been investigated in applications to macroscopic superposition
of quantum states\cite{time_dep_nonlinearity}.

It is well assessed that in the case of unitary dynamics, without
any losses, an anharmonic oscillator leads to sub-Poissonian
statistics of oscillatory excitation number, quadratic squeezing
and superposition of macroscopically distinguishable coherent
states. For dissipative dynamics the important parameter
responsible for production of nonclassical states via ${\chi}(3)$
materials is the ratio between nonlinearity and damping.
Therefore, the practical realization of such quantum effects
requires a high nonlinearity with respect to dissipation. In this
direction the largest nonlinear interaction was proposed in many
papers, particularly, in terms of electromagnetically induced
transparency\cite{EM} and by using the Purcell effect\cite{P},
and in cavity QED\cite{R}. The significant nonlinearity has also
been observed for nanomechanical resonators\cite{N}. These
methods can lead to ${\chi}(3)$ nonlinearity of several orders of
magnitude higher than natural optical self-Kerr interactions.
Note, that high ${\chi}(3)$ nonlinear oscillators generate also a
lot of interest recently due to their applications in areas of
quantum computing\cite{C}.

In the case of nonlinear dissipative ${\chi}(3)$ interaction
stimulated by coherent driving force, the time evolution cannot be
solved analytically for arbitrary evolution times and suitable
numerical methods have to be used. Nevertheless, with dissipation
included a driven AHO model has been solved exactly in the
steady-state regime in terms of the Fokker-Planck equation in the
complex P representation\cite{I}. Analogous solution has been
obtained for a combined driven parametric oscillator with Kerr
nonlinearity\cite{II}. The Wigner functions for both these models
have been obtained using these solutions\cite{I, III}.

The investigation of quantum dynamics of a driven dissipative
nonlinear oscillator for non-stationary cases is much more
complicated and only a few papers have been done in this field up
to now. More recently, the quantum version of dissipative AHO or
the Duffing oscillator with time-modulated driving force has been
studied in the series of the papers\cite{SR-X1} in the context of
a stochastic resonance\cite{SR}, quantum-to-classical transition
and investigation of quantum dissipative chaos\cite{X2, X1}.

For the constant parameters $\chi(t)=\chi$ and $f(t)=f$ the
equations (\ref{eq:master_equatin}) and (\ref{eq:peturbation})
describe the model of a driven dissipative AHO that was introduced
long ago in quantum optics to describe bistability due to a Kerr
nonlinear medium\cite{drumm}. For the case of time-dependent
parameters $\chi(t)$ and $f(t)$ the dynamics of the AHO exhibits a
rich phase-space structure, including regimes of regular, bistable
and chaotic motion. We perform our calculations for regular and
chaotic regimes concerning two models of time-modulated AHO
corresponding to two physical situations: (i) $\chi=\chi(t)=\chi_0
+\chi_1 sin(\Omega t)$ and $f(t)=const=f$; (ii)
$f=f(t)=f_0+f_1\exp(\delta t)$ and $\chi(t)=const=\chi$ with
$\delta \ll \omega$ and $\Omega \ll\omega$ are the modulation
frequencies.

We analyze the master equation numerically using quantum state
diffusion method (QSD)\cite{QSD}. According to this method, the
reduced density operator is calculated as an ensemble mean
\begin{equation}
\rho(t)= M(|\psi_{\xi}(t)\rangle\langle\psi_{\xi}(t)|)= \lim_
{N\rightarrow\infty}\frac{1}{N}\sum_{\xi}^{N}|\psi_{\xi}(t)\rangle\langle\psi_{\xi}(t)|
\end{equation}
over the stochastic pure states $|\psi_{\xi}(t)\rangle$ describing
evolution along a quantum trajectory. The stochastic equation for
the state $|\psi_{\xi}(t)\rangle$ involves both Hamiltonian
described by Eq. (\ref{eq:peturbation}) and the Linblad operators
described by Eq. (\ref{eq:linbland}) and reads as:
\begin{eqnarray}\label{eq.qsd}
|d\Psi_{\xi}\rangle=-\frac{i}{\hbar}H|\Psi_{\xi}\rangle dt-
\frac{1}{2}\sum_{i=1,2}(L_{i}^{+}L_{i}-2<L_{i}^{+}>L_{i}+\nonumber~~~~~~~~~\\
<L_{i}><L_{i}^{+}>)|\Psi_{\xi}\rangle dt+
\sum_{i=1,2}(L_{i}-<L_{i}>)|\Psi_{\xi}\rangle d\xi ,
\end{eqnarray}
where $\xi$ is the generated complex gaussian noise that satisfies
the following conditions:
\begin{equation}
M(d\xi_{i})=0, M(d\xi_{i}d\xi_{j})=0,
M(d\xi_{i}d\xi_{j}^{*})=\delta_{ij}dt.
\end{equation}
We calculate the density
operator using an expansion of the state vector
$|\psi_{\xi}\rangle$ in a truncated basis of Fock's number states
of a harmonic oscillator
\begin{equation}
|\psi_{\xi}(t)\rangle=\sum_{n}a_{n}^{\xi}(t)|n\rangle.
\end{equation}

\section{Wigner function and Poincar$\acute{e}$ section}
In this section we shortly examine correspondence between the
Poincar$\acute{e}$ section and the Wigner function of the
oscillatory mode considering the model (ii) with time-dependent
driving amplitude \cite{X2, X1}. In semiclassical treatment a
chaotic operational regime is analyzed on phase space of
dimensionless position and momentum $x=Re(\alpha)$ and
$y=Im(\alpha)$, where $\alpha=<a>$ is the oscillatory complex
amplitude (see, Eq.(\ref {eq:alpha_force})). Choosing $x_0$ and
$y_0$ as an arbitrary initial phase-space point of the system at
the time $t_0$, we define a constant phase map in the plane by the
sequence of points  at $t_n=t_0+(2\pi/\delta)n$, where
$n=0,1,2...$. This means that for any $t=t_n$ the system is at one
of the points of the Poincar$\acute{e}$ section.

\begin{figure}
\includegraphics[width=4.5in] {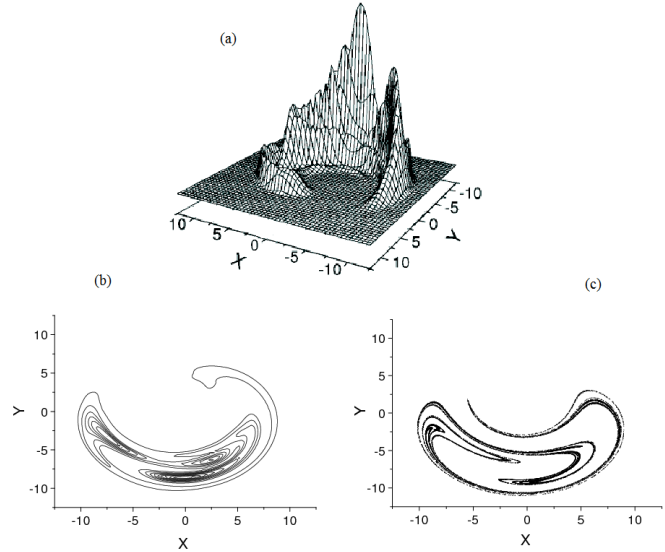}
\caption{The Wigner function (a) and its contour plot (b) averaged over 3000
trajectories. The Poincar$\acute{e}$ section (c) (20000 points plotted at times of the
constant phase for the same parameters). In both cases, the dimensionless
parameters are in the range of chaos, i.e. $\chi/\gamma=0.1$, $\Delta/\gamma=-15$, $f_0/\gamma=f_1/\gamma=27$, $\delta/\gamma=5$.}
\label{fig.wigner_poincare}
\end{figure}

The analysis show that for time scales exceeding the damping time,
$t \gg \gamma^{-1}$, the asymptotic dynamics of the system is
regular in the limits of small and large values of the modulation
frequency, i.e., $\delta \gg \gamma$, $\delta \ll \gamma$ and
small or large amplitude of driving field $f_{1} \ll f_{0}$,
$f_{1} \gg f_{0}$. Fig. \ref{fig.wigner_poincare}(c) shows the
results of numerical calculations of the classical maps, for the
parameters chosen in the range of chaos, i.e. $f_0 \approx f_1$
and $\delta \geq \gamma$. As we see, the figure clearly indicates
the classical strange attractor with fractal structure that is
typical for a chaotic dynamics.

It should be mentioned, that the Wigner function is one of the
quantities that allows to observe chaos in quantum theory. The
nonstationary Wigner function is written as for the state vector $
|\psi(t)\rangle$.

\begin{equation} \label{eq.Wigner_alpha}
W(\alpha, t)=\frac{2}{\pi^{2}}\exp(-2|\alpha|^{2})\int d^2\beta
\langle -\beta|\psi(t)\rangle\langle \psi(t)|\beta\rangle
\exp{(-2(\beta\alpha^{*}-\beta^{*}\alpha))}
\end{equation}

We apply the QSD to determine Wigner functions for the quantum
states of a driven anharmonic oscillator during time evolution.
For this, we use the well-known expression for the Wigner function
in terms of the matrix elements $\rho_{nm}=\langle
n|\rho|m\rangle$ of the density operator in the Fock state
representation:
\begin{equation}\label{eq.WigRT}
W(r, \theta)=\sum_{n,m}\rho_{nm}(t)W_{mn}(r,\theta).
\end{equation}
Here: $(r,\theta)$ are the polar coordinates in the complex
phase-space plane, $x=r\cos\theta$, $y=r\sin\theta$, while the
coefficients $W_{mn}(r,\theta)$ are the Fourier transform of
matrix elements of the Wigner characteristic function:

\begin{equation}
\label{eq.Wigf}
W_{mn}\left(r,\theta \right) = \left\{
\begin{array}{rcl}
\frac{2}{\pi}\left(-1\right)^{n}\sqrt{\frac{n!}{m!}}\exp{i(m-n)\theta}(2r)^{m-n}\\exp{(-2r^2)}L_{n}^{m-n}(4r^2),~ m\geq n\\
\frac{2}{\pi}\left(-1\right)^{m}\sqrt{\frac{m!}{n!}}\exp{i(m-n)\theta}(2r)^{n-m}\\exp{(-2r^2)}L_{m}^{n-m}(4r^2),~
n\geq m .
\end{array},
\right.
\end{equation}

It is remarkable that there is a correspondence between contour
plots of the Wigner function  and the Poincar$\acute{e}$ section.
This point is illustrated on the Fig. \ref{fig.wigner_poincare},
where the Wigner function Fig. \ref{fig.wigner_poincare}(a), its
contour plot, Fig. \ref{fig.wigner_poincare}(b) and
Poincar$\acute{e}$ section, Fig. \ref{fig.wigner_poincare}(c) are
presented for the same parameters. Note, that the Wigner function
is a quasidistribution in phase space averaging an ensemble of
quantum trajectories obtained for a define time moment while
Poincar$\acute{e}$ section is the distribution for time intervals:
it is constructed by fixing points in phase space at a sequence of
periodic moments. What we can conclude from their correspondence
is the fact, that the quantum distribution in phase space
corresponds on the form to the semiclassical distribution but for
big numbers of time intervals. As it is seen, the
Poincar$\acute{e}$ section has fine fractal structure while Wigner
function contour plot has not. This is due to the Hiesenberg
uncertainty relations which prevent sub-Plank structures in phase
space.

\section{Scaling invariance in dissipative chaos}
\begin{figure}
\begin{center}
\psfig{file=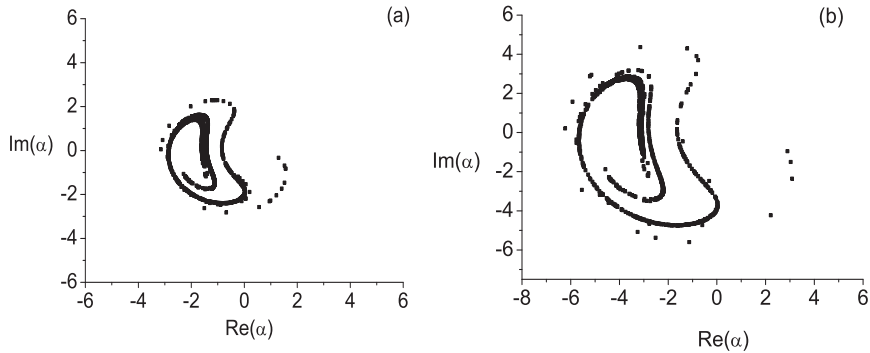,width=4.5in}
\end{center}
\caption{The Poincar$\acute{e}$ section for the following
parameters: $\Delta/\gamma=-15$, $\chi/\gamma=2$,
$f_0/\gamma=5.8$, $f_1/\gamma=4.9$, $\delta/\gamma=2$,
\ref{fig.scaling}((a)); for the scaled $\lambda=2$ parameters
\ref{fig.scaling}((b)). The Poincar$\acute{e}$ sections are
generated for time evaluation 5000 force periods.}
\label{fig.scaling}
\end{figure}

In this section the scaling invariance for the case of chaotic
dynamics is considered on the base of anharmonic oscillator with
time-dependent driven force. In the classical limit the system is
described by the following equation of motion for the
dimensionless amplitude:
\begin{equation}\label{eq:alpha_force}
\frac{d\alpha}{dt}=-\frac{\gamma}{2}\alpha-i(\Delta
+\chi(1+2|\alpha|^{2}))\alpha-i(f_0+f_1\exp(-\delta t)).
\end{equation}
This equation is invariant for the scaling transformation of
complex amplitude $\alpha=\lambda\alpha$ if the other parameters
transforms like:
$\Delta\rightarrow\Delta^{'}=\Delta+\chi(1-1/\lambda^{2})$, $
\chi\rightarrow\chi^{'}=\chi/\lambda^{2}$, $ f\rightarrow
f^{'}=\lambda f$, $\gamma\rightarrow \gamma^{'}=\gamma$. This
scaling property of the classical equation for the chaotic
dynamics leads to the symmetry of strange attractors: they have
the same form in the phase space and differ from each other only
in scale. This fact is demonstrated in the Fig. \ref{fig.scaling}.

Strictly speaking the quantum system does not obey the same
scaling invariance as classical one. An analysis of scaling
invariance from the point of view of quantum-statistical theory
has been performed\cite{X3}. It was shown that such parameter
scaling occurs for wider ranges of the parameters, but for not
large values of the parameter $\chi/\gamma$, where system displays
strong quantum properties.

Now we use the scaling arguments considering stochastic dynamic of
a single trajectory in this way: to investigate chaos we use
Poincar$\acute{e}$ section based on evolution of a single quantum
trajectory, Eq.\ref{eq.qsd}. The Poincar$\acute{e}$ section is
obtained by recording $(Re(\alpha_{\xi}), Im(\alpha_{\xi}))$ at
time intervals of $2\pi/\delta$, where $\xi$ indicates the
stochastic variable and $\alpha_{\xi}$ is obtained from Eq.
\ref{eq.qsd}. Our goal is to analyze the scaling invariance on the
base of this quantity. On the other side we present results that
relies on quantum-classical correspondence.

\begin{figure}
\begin{center}
\psfig{file=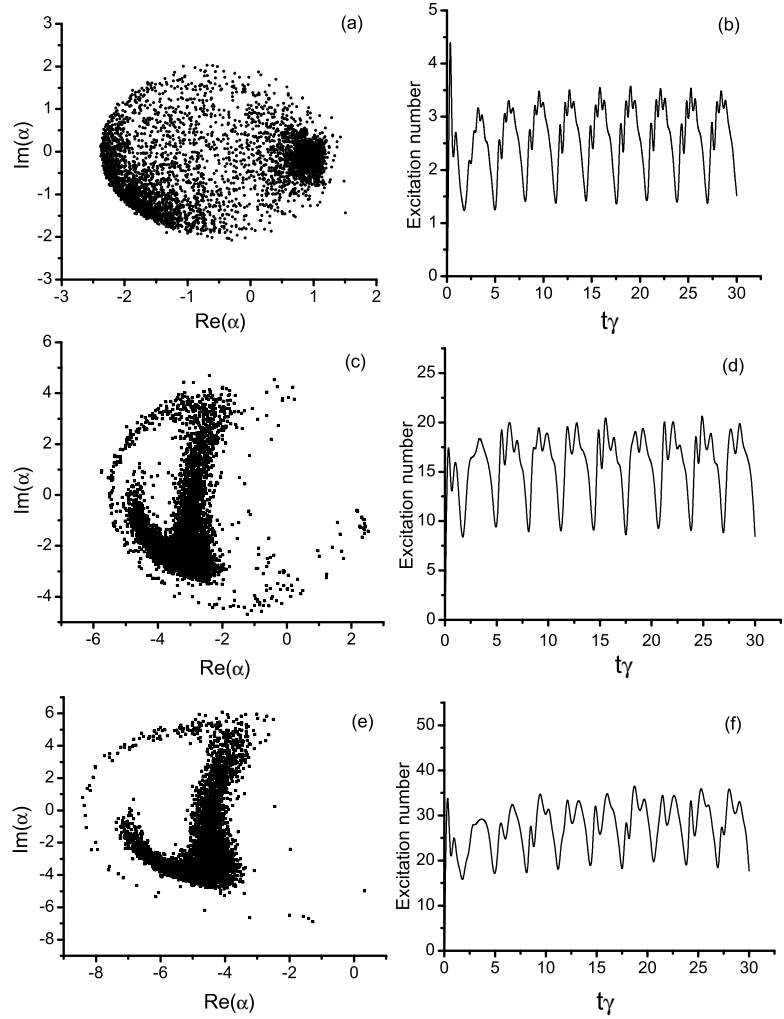,width=4.5in}
\end{center}
\caption{The Poincar$\acute{e}$ section calculated on the base of
a single trajectory for the following parameters:
$\Delta/\gamma=-15$, $\chi/\gamma=2$, $f_0/\gamma=5.8$,
$f_1/\gamma=4.8$, $\delta/\gamma=2$, (Fig. \ref{fig.poincare} (a));
the excitation number for the same parameters (Fig.
\ref{fig.poincare}(b)). Figs. \ref{fig.poincare}(c) and
\ref{fig.poincare}(d) corresponds to scaled $\lambda=2$
parameters. Figs. \ref{fig.poincare}(e) and \ref{fig.poincare}(f)
corresponds to scaled $\lambda=3$ parameters. The
Poincar$\acute{e}$ sections are generated for time evaluation 5000
force periods.}\label{fig.poincare}
\end{figure}

In this direction, we consider the specific parameters for which
DAO is in the vicinity of chaotic behavior determined classically.
That means, if the parameters are slightly tuned in this range the
transition from chaotic to regular dynamics might be realized. As
we have realized above the dynamics of the system is chaotic in
the range of parameters $f_0\simeq f_1$ and $\delta\geq\gamma$.
Particularly, it is demonstrated on the base of the semiclassical
Poincar$\acute{e}$ section the system exhibits chaotic dynamics
for $f_0/\gamma=f_1/\gamma=5.8$, $\Delta/\gamma=-15$ and
$\delta/\gamma=2$ in classical treatment. As analysis shows for
the considered parameters the system dynamics continues to be
chaotic till $f_1/\gamma=4.9$(Fig.\ref{fig.scaling}), while
chaotic dynamics becomes regular for $f_1/\gamma=4.8$. Thus, we
examine Poincar$\acute{e}$ section in quantum trajectory started
from the regular regime, for the parameter $f_1/\gamma=4.8$. The
results of calculations are presented in Fig. \ref{fig.poincare}
for the parameters: $\Delta/\gamma=-15$, $\chi/\gamma=2$,
$f_0/\gamma=5.8$, $\delta/\gamma=2$, $f_1/\gamma=4.8$. As we see,
for the scaled parameter $\lambda=1$ the Poincar$\acute{e}$
section is distributed stochastically but approximately exhibits
localization in two ranges (Fig. \ref{fig.poincare}(a)); its shape
does not coincide with Poincar$\acute{e}$ section
(Fig.\ref{fig.scaling}(c)). For the case of $\lambda=2$, points of
the Poincar$\acute{e}$ section on a trajectory are distributed
chaotically, however, unlike to the previous case $\lambda=1$,
here an correspondence between both shapes take place. This
likeness become more obvious for the case of $\lambda=3$ (Fig.
\ref{fig.poincare}(e)).

It is important to note that in the semi-classical treatment all
cases $\lambda=1,2,3$ correspond to regular regime as it is
expected due to scaling invariance. Another situation is realized
in the quantum treatment. For $\lambda=1$ the system seems to be
in regular regime. This statement is also confirmed by calculation
of the Wigner function. The calculations shows that for such
parameters
 the Wigner function has two-peak structure indicating
that the regime of bistability is realized. If we increase the
scaling parameter the resulted Poincar$\acute{e}$ section's shapes
in a quantum trajectory in their forms are coincided with
Poincar${\acute{e}}$ sections in semiclassical treatment. So we
found the parameters for which the system in classical treatment
has regular dynamic while in quantum treatment its dynamic is
chaotic.

The results on mean excitation numbers are presented on Fig.
\ref{fig.poincare}(b, d, f). As we see, for $\lambda=1$, the
oscillatory excitation numbers varies from 1 to 4 and thus the
system is in deep quantum regime. The level of quantum noise is
comparatively sufficient and chaos cannot exhibit itself on
Poincar$\acute{e}$ section. But for scaled parameters
$\lambda=2,3$ the excitation numbers increase and varies up to 40
(Fig. \ref{fig.poincare}(d,f)). For these cases the ranges of
variation are much enough to exhibit a chaotic type structures.

Note, the usefulness of scaling procedure. It allows us to analyze
classically analogy regimes for various groups of the parameters.
In particular, increasing scaling parameter of the system we can
investigate quantum-classical correspondence.

\section{Quantum interference assisted by chaos}
It is well assessed that in the case of unitary dynamics, without
any losses, an anharmonic oscillator leads to sub-Poissonian
statistics of oscillatory excitation number, quadratic squeezing,
and superposition of macroscopically distinguishable coherent
states.The dissipation and decoherence lead usually to losses of
these effects. In this section, we have numerically studied the
phenomena at the overlap of chaos, dissipation, and quantum
effects for the time-dependent nonlinear model. It was recently
shown\cite{X1, X3, mer} that physical systems based on this model
have a potential for generation of high-degree sub-Poissonian
light as well as for the observation of quantum-statistical
effects and quantum interference that accompanied by chaotic
dynamics. Here we concentrate on studies of quantum interference.
We have pointed out that the time modulation of the oscillatory
parameters, which are the strength of third-order nonlinearity or
the amplitude of the driving force, leads to formation of the
quantum-interference patterns in phase space in over transient
regimes, for the definite time intervals exceeding the transient
dissipation time.

It is well known that the phase-space Wigner distribution function
can simply visualize nonclassical effects including
quantum-interference. For example, a signature of quantum
interference is exhibited in the Wigner function by non-positive
values. In this section the numerical results of the nonstationary
Wigner functions in chaotic regimes of AHO are presented and
discussed.

It should be noted that the most of investigations of the quantum
distributions of oscillatory states, including also modes of
radiations, have been made for the steady-state situations. The
simplicity of Kerr nonlinearity allows to determine the Wigner
function of the quantum state under time evolution due to
interaction. In this sense, we note the main peculiarity of our
paper in comparison with above noted important inputs. In this
paper, we calculate the Wigner functions in an over transient
regime, $t\gg\gamma^{-1}$, of the dissipative dynamics, however,
we consider time-dependent effects which appear due to the
time-modulation of the oscillatory parameters.

\begin{figure}
\begin{center}
\psfig{file=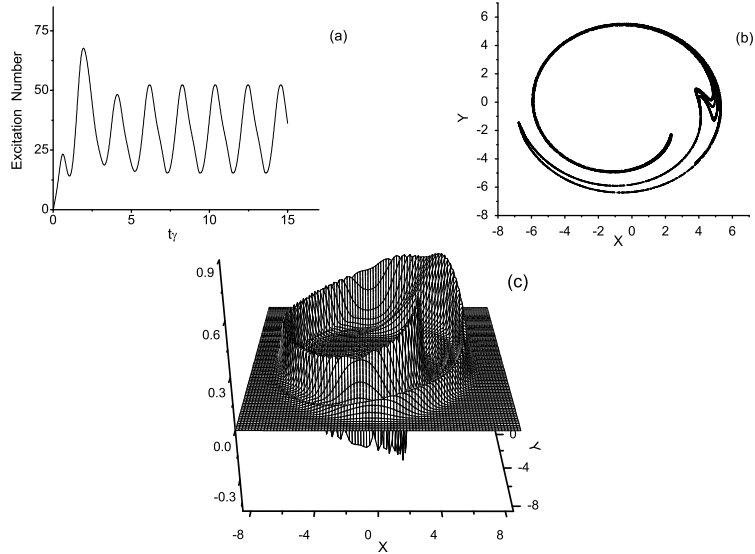,width=4.5in}
\end{center}
\caption{The mean excitation number (a); the $Poincar\acute{e}$
section (b) ($ \approx 20000$ points) for the dimensionless
complex amplitude $\alpha$, plotted at times of the constant phase
$\delta t_{k}= 2 \pi k$ (k = 0,1,2...), when the maximal
interference pattern on the Wigner function (c) is realized, for
the case of time-modulated nonlinearity $\chi(t)$. The parameters
are in the range of chaos:
$\Delta/\gamma=5,\chi(t)/\gamma=0.2(1+0.75\sin(\Omega
t)), \Omega/\gamma=3, f/\gamma=10$.} \label{fig.chaosinterf}
\end{figure}

Below we investigate the Wigner functions for time-modulated
nonlinearity. Note that the dynamics of the system with
time-modulated nonlinear strength AHO is chaotic in the ranges:
$\delta\geq\gamma$ and $\chi_{0}\simeq \chi_{1}$ and for negative
detuning. As shows analysis, controlling transition from the
regular to chaotic dynamics can be realized through the
intermediate ranges of bistability by varying  the strength
$\chi_{1}$ of the modulation processes in the ranges from
$\chi_{1}\ll\chi_{0}$ to $\chi_{1}\leq\chi_{0}$. The results of
the ensemble averaged numerical calculation of the mean excitation
number, the Poincaré section and the Wigner function are shown on
the Figs. \ref{fig.chaosinterf} respectively. The mean excitation
number of the driven AHO versus dimensionless time is depicted in
Fig. \ref{fig.chaosinterf} (a). In contrast to the semiclassical
result its quantum ensemble counterpart (see, Fig.
\ref{fig.chaosinterf}(a)) has clear regular periodic behavior for
time intervals exceeding the characteristic dissipation time, due
to ensemble averaging. The Fig. \ref{fig.chaosinterf}(b) clearly
indicates the classical strong attractors with fractal structure
that are typical for a chaotic $Poincar\acute{e}$ section. Thus,
the Wigner function (Fig. \ref{fig.chaosinterf}(a)) reflects the
chaotic dynamics, its contour plots in the $(x,y)$ plane are
similar to the $Poincar\acute{e}$ section. However, the Wigner
functions have regions of negative values for the definite time
intervals. The example depicted on Fig. \ref{fig.chaosinterf}(a)
corresponds to time intervals $\gamma t_{k}= 6 + \frac{2 \pi
k}{\delta}\gamma$ $(k=0,1,2...)$, for which the mean excitation
number reaches a macroscopic level, i.e., $n=52$. The interference
pattern is destroyed as the time modulation is decreased. Indeed,
it is shown\cite{I, II} that for oscillatory mode of driven
anharmonic oscillator including dissipation the Wigner function is
positive in all phase space.

\section{Conclusion}

In summary, we have discussed dissipative chaos answering what is
the counterpart of the semiclassical Poincar$\acute{e}$ section in
quantum treatment. We have presented a type of nonstationary
systems showing chaotic dynamics and intrinsically quantum
properties which are modelled by a driven dissipative anharmonic
oscillator with time-dependent parameters.  The connection between
quantum and classical treatments of chaos has been realized by
means of a comparison between strange attractors on the
semiclassical Poincar$\acute{e}$  section, the shapes of the
Poincar$\acute{e}$ section on a single quantum trajectory and the
contour plots of the Wigner functions. We have analysis the
borders of validity of scaling invariance for quantum dissipative
chaos and we have demonstrated realization of long-lived quantum
interference assisted by chaotic dynamics for over transient
regime and for the macroscopic level of oscillatory excitation
numbers.

\section*{Acknowledgments}
This work was supported by NFSAT/CRDF grant UCEP-02/07 and ISTC
grants A-1517 and A-1606.


\begin{thebibliography}{9}

\bibitem{quant_chaos} M. C. Gutzwiller, Chaos in Classical and Quantum Systems
(Springer-Verlag, Berlin, 1990); Quantum Chaos, Quantum Measurement, edited by P. Citanovic, I. Percival, and A.
Wizba (Kluwer, Dordrecht, 1992); K. Nakamura, Quantum Chaos, A New Paradigm of Nonlinear Dynamics, Vol. 3 of Cambridge Nonlinear Science Series (Cambridge University Press Cambridge, 1993); Quantum Chaos, edited by G. Casati and B. Chirikov (Cambridge University Press, Cambridge,
1995); F. Haake, Quantum Signatures of Chaos (Springer-Verlag, Berlin, 2000).

\bibitem{back} E. Ott, M. Antonsen, Jr., and J.D. Hanson, Phys. Rev. Lett. {\bf 53},
2187 (1984); T. Dittrich and R. Graham, Ann. Phys. (N.Y.) {\bf 200}, 363 1990).

\bibitem{lyap} A. Peres, in Quantum Chaos: Proceedings of the Adriatico
Research Conference on Quantum Chaos, edited by H. A. Cerdeira
et al. (World Scientific, Singapore 1991); A. Peres, Quantum Theory: Concepts and Methods (Kluwer, Dordrecht,
1993); R. Schack and C.M. Caves, Phys. Rev. Lett. {\bf 71}, 525 (1993).

\bibitem{zur} W. H. Zurek and J. P. Paz, Phys. Rev. Lett. {\bf 72}, 2508 (1994); {\bf 75},
351 (1995); S. Habib, K. Shizume, and W. H. Zurek, ibid. {\bf 80},
4361 (1998); J. Gong and P. Brumer, Phys. Rev. E {\bf 60}, 1643 (1999).

\bibitem{zur_2} W. H. Zurek, Nature (London) {\bf 412}, 712 (2001).

\bibitem{QSD} N. Gisin and I. C. Percival, J. Phys. A {\bf 25}, 5677 (1992); {\bf 26}, 2233 (1993); {\bf 26}, 2245 (1993); I. C. Percival, Quantum State Diffusion(Cambridge Unoversity Press, Campridge, (2000).

\bibitem{spil} T. P. Spiller and J.F. Ralph Phys. Lett. A {\bf 194}, 235 (1994).

\bibitem{perc} T.A. Brun, I.C. Percival, and R. Schack, J. Phys. A {\bf 29},
2077 (1996); T. Bhattacharya, S. Habib, and K. Jacobs, Phys. Rev. Lett. {\bf 85}, 4852 (2000).

\bibitem{SR} H.\ H.\ Adamyan,\ S.\ B.\ Manvelyan,\ and G.\ Yu.\ Kryuchkyan\, Phys.\ Rev.\ A {\bf
63}, 022102 (2001).

\bibitem{X2} H.\ H.\ Adamyan,\ S.\ B.\ Manvelyan,\ and G.\ Yu.\ Kryuchkyan\,
Phys.\ Rev.\ E{\bf 64}, 046219 (2001).

\bibitem{X1} G.\ Yu.\ Kryuchkyan\ and S.\ B.\ Manvelyan,\ Phys.\ Rev.\ Lett.\ {\bf
88}, 094101 (2002);

\bibitem{X3} G.\ Yu.\ Kryuchkyan\ and S.\ B.\ Manvelyan, Phys. Rev. A{\bf 68}, 013823 (2003).

\bibitem{mer} T. G. Gevorgyan, A, R. Shahinyan, G. Yu. Kryuchkyan Phys. Rev. A. {\bf 79}, 062905, (2009).

\bibitem{N} I.\ Kozinsky, H.\ W.\ Ch.\ Postma,\ O.\ Kogan, A.\ Husain and M.\ L.\ Roukes,\ Phys.\ Rev.\ Lett. {\bf 99},
207201 (2007).

\bibitem{amp} H. W. Ch. Postma, I. Kozinsky, A.
Hussian, and M. L. Roukes, Appl. Phys. Lett. {\bf 86}, 223105
(2005).

\bibitem{nano} X.\ M.\ H.\ Huang, C.\ A.\ Zorman, M.\ Mehregany, and M.\ L.\ Roukes, Nature (London) {\bf
412}, 496 (2003); K. C. Schwab and M. L. Roukes, Phys. Today {\bf
58}, N7, 36 (2005).

\bibitem{nano_1}A. N. Cleland and M. R. Geller, J.
App. Phys. {\bf 92}, 2758 (2002); A. N. Cleland and M. R. Geller,
Phys. Rev. Lett. {\bf 93}, 070501 (2004).

\bibitem{Qnr1} I. Katz, A. Retzker, R. Straub, and R. Lifshitz, Phys. Rev. Lett. {\bf 99} 040404
(2007).

\bibitem{Qnr2} E. Buks, E. Segev, S. Zaitsev, B. Abdon, and M. P. Blencowe, Europhys. Lett., {\bf 81} 10001
(2008).

\bibitem{Qnr3} R. Almog, S. Zaitsev, O. Shtempluck, and E. Buks, App. Phys. Lett. {\bf 90}, 013508 (2007); J. S. Aldridge and A. H. Cleland, Phys. Rev. Lett {\bf 94}, 156403
(2005).

\bibitem{A} R.\ Almog, S.\ Zaitsev, O.\ Shtempluck, and E. Buks, Phys.\ Rev.\ Lett.\ {\bf
98}, 078103 (2007).

\bibitem{duffing_force} E.\ Babourine-Brooks, A.\ Doherty, G.\ J.\ Milburn, quant-ph/ 0804.3618v1, (2008).

\bibitem{devices} M. P. Blencowe, Phys. Rep. {\bf 395} 159 (2004); T. J. Kippling and K. J. Vahala, Opt. Expr. {\bf 15},17172 (2007);
M. Aspelmeyer and K. Schwab, New J. Phys. {\bf 10}, 095001 (2008).

\bibitem{a} A. E. Kaplan, Phys.\ Rev.\ Lett.\ {\bf 48}, 138
(1982).

\bibitem{b} D. Enzer and G. Gabrielse, Phys.\ Rev.\ Lett.\ {\bf 78}, 1211 (1997); G. Gabrielse, H. G. Dehmelt, and W. Kells, ibid. {\bf 54}, 537
(1985).

\bibitem{c} M. Rigo, G. Alber, F. Mota-Furtado, and P. F. OMahony, Phys. Rev. A{\bf 55}, 1665 (1997); A{\bf 58}, 478
(1998).

\bibitem{trapped_ion} F. Diedrich, J. C. Berquist, W. M. Itano, and D. J. Wineland, Phys. Lett. {\bf 62}, 403 (1989); D. J. Wineland, C. Monroe, W. M. Itano, D. Liebfred, B. E. King and D. M. Mekhot, J. Res. Natn. Inst. Standt. Technol. {\bf 103}, 259 (1998).

\bibitem{time_dep_nonlinearity} S.\ Mancini and P.\ Tombesi, Phys. Rev. {\bf A52}, 2475 (1995).

\bibitem{EM} A. Imamoglu, H. Schmidt, G. Woods, and M. Deutsch, Phys. Rev. Lett. {\bf 79}, 1467 (1997); M. J. Werner and A. Imamoglu,
Phys. Rev. A {\bf 61} 011801(R) (1999); M.\ Fleischhauer\, A.\
Imamoglu, and J.\ P.\ Marangos,\ Phys.\ Mod.\ Phys. {\bf 77}, 633
(2005).

\bibitem{P} P.\ Bermel, A. Rodriguez, J. D. Joannopoulos, and M. Soljacic, Phys.\ Rev.\ Lett. {\bf 99}, 053601
(2007).

\bibitem{R} F. G. S. L. Brandao, M. Hertmann, and M. B. Plenio, New J.\ Phys.\
{\bf 10}, 043010 (2008).

\bibitem{C} Q.\ A. Turchette , C. J. Hood, W. Lange, H. Mabuchi, and H. J. Kimble, Phys.\ Rev.\ Lett. {\bf 75}, 4710
(1995); K.\ Nemoto, W. J. Munro, Phys.\ Rev.\ Lett. {\bf 93},
250502 (2004); W. J. Munro, K. Nemoto and T. P. Spiller, New J.
Phys. {\bf 7}, 137 (2005); J. Lee, M. Paternostro, C. Ogden, Y. W.
Cheong, S. Bose and M. S. Kim, New J. Phys. {\bf 8}, 23 (2006).

\bibitem{I} K.\ V.\ Kheruntsyan\, J.\ Opt.\ B:\ Quantum Semiclass.\ Opt. {\bf
1}, 225 (1999).

\bibitem{II} G.\ Yu.\ Kryuchkyan\ and K.\ V.\ Kheruntsyan\, Opt.\ Commun\ {\bf
120}, 132 (1996).

\bibitem{III} K.\ V.\ Kheruntsyan\, D.\ S.\ Krahmer, G.\ Yu.\ Kryuchkyan\ and K.\ G.\ Petrossian,\ Opt.\ Commun\ {\bf
139}, 157 (1997).

\bibitem{drumm} P. D. Drummond and D. F. Walls, Phys. Rev. A {\bf 23} 2563 (1981).



\end{thebibliography}
\end{document}